\documentclass{emulateapj}
\newcommand{\beq}{\begin{equation}}
\newcommand{\eeq}{\end{equation}}
\def\etal{{\em et al. }}

\begin{document}


\title{Cosmic acceleration vs axion-photon mixing}

\shorttitle{Cosmic acceleration vs axion-photon mixing}

\author{Bruce A. Bassett}
\affil{Department of Physics, Kyoto University, Kyoto, Japan\&Institute of Cosmology and Gravitation, University of Portsmouth, Portsmouth~PO1~2EG, UK}
\email{Bruce.Bassett@port.ac.uk}
\and
\author{Martin Kunz}
\affil{Astronomy Centre, University of Sussex, Falmer, BN1 9QH, UK}
\email{M.Kunz@sussex.ac.uk}      


\begin{abstract}
Axion-photon mixing has been proposed as an alternative to acceleration as the explanation for supernovae dimming. We point out that the loss of photons due to this mixing will induce a strong asymmetry between the luminosity, $d_L(z)$, and angular-diameter distance, $d_A(z)$, since the latter is unaffected by mixing. In a first search for such an asymmetry we introduce a dimensionless mixing amplitude $\lambda$ so that $\lambda=0$ if no photons are lost and $\lambda\approx1$ if axion-photon mixing occurs. The best-fit to  SNIa and radio galaxy data is  $\lambda = -0.3^{+0.6}_{-0.4}$ (95\% CL), corresponding to an unphysical, negative, mixing length. This same argument limits the attenuation of light from supernovae due to dust. We show that future $d_L$ and $d_A$ data from SNAP and galaxy surveys such as DEEP2 and KAOS will detect or rule out mixing at more than 5$\sigma$, almost independently of the dark energy dynamics. Finally we discuss the constraints from the near maximal polarisation of the gamma-ray burst (GRB) GRB021206. Since mixing reduces the polarisation of distant sources, future observations of high redshift GRBs may provide orthogonal constraints on axion-photon mixing and related scenarios. 
\end{abstract}

\section{Introduction}

The reciprocity relation is a wonderfully powerful result valid for any metric theory of gravity where photons travel on null geodesics, as long as photon number is conserved (Etherington 1933; Ellis 1971).  It ensures that the luminosity distance, $d_L(z)$ is exactly the same as the angular-diameter distance, $d_A(z)$, up to a factor of $(1 + z)^2$. In this paper we turn the reciprocity relation around and use it to probe alternatives to cosmic acceleration. 

The accumulating evidence for recent cosmic acceleration (Barris {\em et al} 2003; Knop {\em et al}, 2003) leaves us with the familiar coincidence problem - why do we live at such a special time? 
An attractive alternative is that the acceleration is a mirage and not a real feature of the dynamics of our Universe. 

Although such a non-accelerating cosmology can be made reasonably compatible with cosmic microwave background (CMB) and large scale structure (LSS) data (Blanchard {\em et al.} 2003), the dilemma then is to explain, without acceleration, the dimming of distant Type Ia supernovae (SNIa) and the observed $\sim 3\sigma$ correlation between the CMB and LSS (Boughn \& Crittenden 2003; Nolta \etal 2003; Scranton \etal 2003; Fosalba \etal 2003). The latter can perhaps be explained by negative spatial curvature, while the dimming of supernovae can be explained by axion-photon mixing (Cs\'aki {\em et al} 2002), meaning that the evidence for acceleration is not yet overwhelming.  

The basic idea of axion-photon mixing is simple. On average, and on large scales relative to the mixing length, $1/3$ of photons in the visible would be lost through conversion to a light axion state, $a$, in the presence of the cosmic magnetic field $\overrightarrow{B}$. This proceeds through the axion interaction term $\frac{a}{4M}\overrightarrow{E}\cdot\overrightarrow{B}$. Cs\'aki {\em et al.} (2002) argue that an axion mass scale of $M \simeq 4\times 10^{11} GeV$ would provide a good fit to SNIa luminosity data as a function of redshift, (quantified in Erlich \& Grojean 2001), without the need for cosmic acceleration, while still being (marginally) consistent with other constraints (see Deffayet {\em et al} 2002; Mortsell {\em et al.} 2002; Christensson and Fairbairn 2003; Mortsell and Goobar, 2003), especially if non-flat FLRW models are considered. 

Intriguingly, axion-photon (AP) mixing, with a similar mass scale, can explain the existence of super-GZK cosmic rays (Cs\'aki {\em et al.} 2003) if the primaries are taken to be photons, since they can travel most of the way as axions, then oscillate back into photons before reaching earth. Axion-photon mixing can therefore provide a simultaneous solution to the super-GZK and coincidence problems and is thus worth further detailed study. 

We make four points about the axion-photon mixing scenario:
\begin{enumerate}
\item Axion-photon mixing should induce a violation of the reciprocity relation and a fundamental disagreement between the dimensionless coordinate distance, $y(z)$, inferred from the luminosity, $d_L(z)$, and angular-diameter distances, $d_A(z)$. 
\item The observed $d_A(z)$ should correspond to a decelerating universe if axion-photon mixing is the source of supernovae dimming. However, current estimates of $d_A(z)$,  favour an accelerating Universe.  
\item Future data from SNAP and KAOS will allow for constraints beyond the 5$\sigma$ level. Tests of number counts from the on-going DEEP2 survey will provide further tight constraints. 
\item Mixing leads to depletion of the polarisation levels of extra-galactic sources (Cs\'aki {\em et al.} 2002; Mortsell and Goobar 2003). The near maximal polarisation seen in the gamma-ray burster (GRB) GRB021206 (Coburn and Boggs, 2003) suggests that GRB's may provide powerful constraints on the mixing scenario when more data is available. \end{enumerate}
In this paper we will assume a flat FLRW model consisting of dust ($\Omega_M$) and dark energy $X$ ($\Omega_X=1-\Omega_M$) with a constant equation of state $w = p_X/\rho_X$. 

\section{The reciprocity relation}

One may define several distances in cosmology. The luminosity distance, $d_L(z)$, estimates distances by comparing the absolute luminosity of an object to its observed/apparent luminosity. The angular-diameter distance, $d_A(z)$, estimates distances based on how the apparent linear size of an object changes with redshift. In metric theories where photons travel on null geodesics and their number is conserved one can show that these two distances are fundamentally related by the reciprocity relation (Schneider {\em et al}, 1992):
\beq
d_L(z) = (1 + z)^2 d_A(z) \,.
\label{recip}
\eeq
When the reciprocity relation holds, the dimensionless coordinate distance $y(z)$ can be estimated from either $d_L(z)$ or $d_A(z)$ via the relations $y(z) = H_0 d_L/(1+z) = H_0 d_A(1 + z)$, where $H_0$ is the current value of the Hubble constant.

In stark contrast the reciprocity relation is not obeyed in the axion-photon mixing scenario, nor indeed in any scenario (such as light attenuation due to dust) which effectively violates photon number conservation. As a result $y(z)$ estimated from $d_L$ and $d_A$ data should disagree since $d_A(z)$ is unaffected but the luminosity distance is modified as $d_L \rightarrow d_L / \sqrt{P_{\gamma\rightarrow \gamma}}$, where $P_{\gamma\rightarrow \gamma}$ is the probability that a photon will reach earth in a photon state and hence be detected. We use the rather good approximation
\beq
P_{\gamma\rightarrow \gamma} = \frac{2}{3}+\frac{1}{3} e^{-l/l_{\rm dec}}
\eeq
where $l$ is the proper distance of the supernovae.
For SNIa at cosmological distances the mixing saturates at $2/3$ (Cs\'aki \etal 2002) and hence supernovae should appear $\sqrt{3/2}$ times further away than they really are, in good agreement with the $d_L(z)$ predicted by the standard best-fit $\Lambda$CDM model with $\Omega_{\Lambda} \simeq 0.7$ and $\Omega_M \simeq 0.3$. The precise value of the decay length $l_{\rm dec}$ depends on the mixing and the galactic magnetic fields. The preferred value of Cs\'aki \etal 2002 is $l_{\rm dec} \approx 1/(2H_0)$. This motivates the use of a dimensionless suppression amplitude 
\beq
\lambda \equiv \left(2 H_0 l_{\rm dec}\right)^{-1}
\eeq
which is zero if no mixing occurs and one in the case of mixing over cosmological distances.

\section{Constraints from current data} 

If axion-photon mixing is to solve the coincidence problem then we should expect the $d_A(z)$ data to fit best to a non-accelerating Universe. This is not the case, however. There are (at least) three independent data sets for $d_A(z)$ that give large best-fit values of $\Omega_{\Lambda}$, consistent with standard $d_L(z)$ best-fits and an accelerating Universe.

Daly \& Guerra (2002) and Daly \& Djorgovski (2003) analysed data for 20 bright 
FRIIb radio galaxies at redshifts between $0.43$ and $1.79$ and, assuming a flat universe, found $\Omega_{M} < 0.5$ and $-2.5 < w < -0.25$ at the $90\%$ confidence level where $w$ is the equation of state of the dominant, non-dust component. Another analysis (Jackson, 2003; Jackson and Dodgson, 1997) of ultra-compact 
radio sources (Gurvitis 1994; Lima and Alcaniz 2002) at $z > 0.5$ found that the best-fit 
flat $\Lambda$CDM model has $\Omega_M = 0.24^{+0.09}_{-0.07}$.
\begin{figure}[tbp]\begin{center}
\includegraphics[width=80mm]{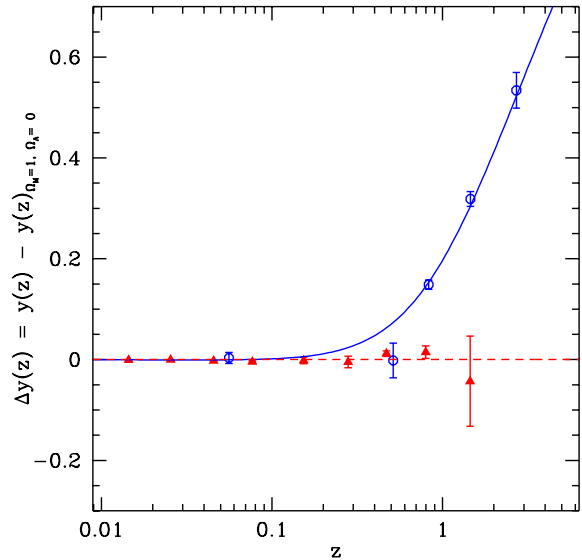} \\
\caption[1dplots]{The residual, $\Delta y(z)$, in the dimensionless coordinate distance relative to an $(\Omega_M,\Omega_{\Lambda})=(1,0)$ universe for redshift-binned SNIa data ($d_L(z)$, solid triangles) ``corrected" for the loss of photons predicted by axion-photon mixing and redshift-binned radio galaxy data ($d_A(z)$, circles). The solid curve is $\Delta y(z)$ for a $\Omega_M = 0.22, \Omega_{\Lambda} = 0.78$ Universe. If mixing were the origin of supernovae dimming we should expect the radio galaxy data to coincide with the best-fit $y(z)$ curve of this corrected SNIa data, near $\Delta y(z)=0$; however the radio galaxy data lies systematically above this curve, favouring an accelerating universe.}
\label{data}
\end{center}
\end{figure}
\begin{figure}[tbp]\begin{center}
\includegraphics[width=80mm]{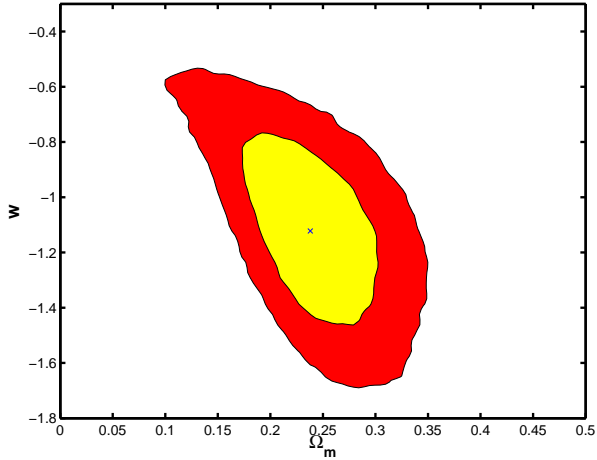} \\
\caption[1dplots]{The 2d likelihood plot for the equation of state $w$ and $\Omega_M$ for the combined SNIa \& RG data sets, with 1 and 2$\sigma$ contours shown. A flat FRW universe was assumed. The combined data clearly favour an accelerating low-density universe.}
\label{2domw}
\end{center}
\end{figure}

Searches for co-moving standard rulers via peaks in the two-point correlation 
function of quasars have also been undertaken. Mamon and Roukema (2002), using a subset of the 
2QZ 2df quasar survey, estimated $\Omega_{\Lambda} = 0.65 \pm 0.35$. Assuming a flat universe they constrain the 
equation of state of the non-dust matter to $w < -0.35$ at $2\sigma$. This approach has been extended 
recently by Outram et al (2003) using the full 2QZ survey, allowing for even stronger results. 
Assuming a flat FLRW model, they find $\Omega_{\Lambda}=0.71^{+0.09}_{-0.17}$ and exclude 
an $\Omega_{\Lambda}=0$ Universe at over 95\% confidence. 

In summary, all current estimates of $d_A(z)$ favour an accelerating Universe, and since they are unaffected by axion-photon mixing, disfavour it as the explanation for the majority of SN-Ia dimming. This point is made visually in figure (\ref{data}) which shows the binned valus of the dimensionless coordinate distance, $y(z)$, of the data sets used in our analysis. The SN-Ia data has been corrected for mixing, and indeed clusters around the non-accelerating EdS model $(\Omega_M,\Omega_{\Lambda}) = (1,0)$. We used the combined data sets of Tonry et al (2003) and Barris et al (2003) to which we added the new supernovae from Knop et al (2003). However, the radio galaxy data clearly prefers an accelerating model. This data is a combination of Daly and Djorgovski (2003), Jackson (2003) and Gurvits (1994). A more detailed description of the data sets can be found in Bassett and Kunz (2003).

In figure (\ref{2domw}) we show the 2d likelihood plot for $\Omega_M$ and $w$ for the combined data sets assuming a flat Universe. 
The overall best-fit  with axion-photon mixing is $\Omega_M = 0.24$, $w = -1.1$ and $\lambda = -0.3$. The one-dimensional (marginalised) 95\% confidence limits on the parameters are $0.15 < \Omega_M < 0.33$, $-1.6 < w < -0.6$ and $-0.7 < \lambda < 0.3$. The values preferred by axion-photon mixing ($w=-1/3$ and $\lambda\approx 1$) are ruled out at well over $3 \sigma$.

Negative values of $\lambda$ correspond to the appearance of photons, instead of absorption. This is not impossible in the axion-photon scenario, e.g. if SN-Ia produce large numbers of axions which then become photons on the way to earth. 
Still, it may be argued that this region of the parameter space is unphysical and should be excluded. In this case the 95\% upper limit (one sided) on $\lambda$ is $0.6$, and $\lambda = 1$ lies at $3 \sigma$. This is no longer sufficient to reliably rule out axion-photon mixing. But the limits on the equation of state remain rather strong, with a 95\% confidence interval of $-1.1 < w < -0.5$ and $w = -1/3$ remains ruled out at over $3 \sigma$. This renders the scenario quite unattractive, as the mixing does not alleviate the coincidence problem asssocitated with cosmic acceleration.

\begin{figure}[tbp]\begin{center}
\includegraphics[width=80mm]{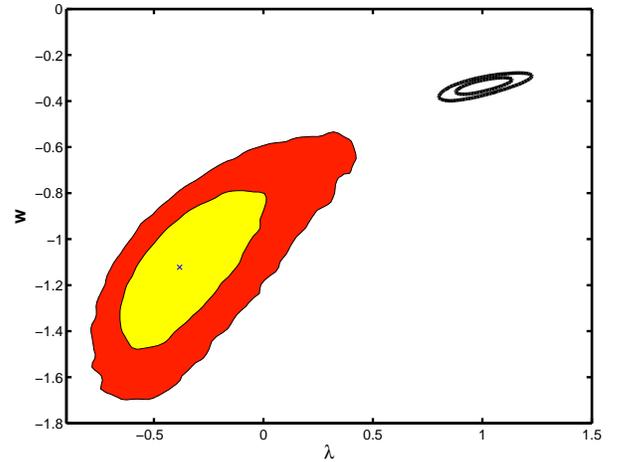} \\
\caption[2dplots]{$\lambda-w$ likelihood  plot for current SNIa and RG data which peaks at $(\lambda,w) = (-0.3,-1.1)$, favouring no mixing ($\lambda=0$) and acceleration. Axion-photon mixing corresponds to $\lambda\approx1$. The likelihood contours for future data, based on a fiducial non-accelerating model with mixing: $(\lambda,w) = (1,-1/3$), are also shown, at the top right.}
\label{2daw}
\end{center}
\end{figure}
\begin{figure}[tbp]\begin{center}
\includegraphics[width=80mm]{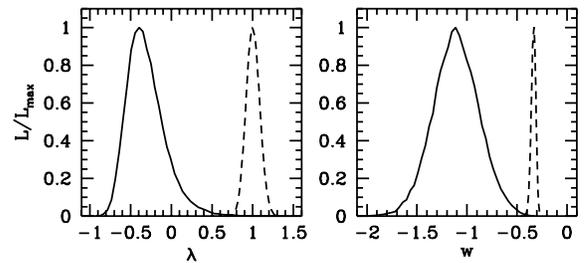}\\
\caption[1dplots]{The 1d likelihood plots for $\lambda$ and $w$ for current data (solid line) and hypothetical future data (dashed curve). $\lambda=0$ corresponds to no photon loss, $\lambda=1$ corresponds to axion-photon mixing. Current data gives $\lambda=-0.3 \pm 0.25~(1\sigma)$. The fiducial model for the future data is flat with $(\lambda,w)=(1,-1/3)$ and $\Omega_M=0.3$.}
\label{1daw}
\end{center}
\end{figure}

\section{Constraints from future data}
Future estimates of $d_L(z)$ from the SNAP satellite\footnote{http://snap.lbl.gov} and $d_A(z)$ from number counts from the DEEP2 survey and from baryon oscillations from the KAOS survey \footnote{http://www.noao.edu/kaos/} will allow estimates of $y_L(z)$ and $y_A(z)$ at the level of a few percent (Aldering {\em et al}, 2002; Linder 2003; Blake \& Glazebrook 2003; Seo \& Eisenstein 2003). To investigate the power of future experiments we used the errors given in the dark energy science case of the KAOS purple book for both the KAOS and SNAP experiments. The central values of the data are chosen to match a model with mixing with an underlying flat FLRW cosmology with $\Omega_M = 0.3$ and $w = -1/3$ which corresponds to the best-fit of Cs\'aki \etal (2002). 

Assuming the auxiliary cosmic parameters (e.g. $\Omega_M$) are well-known from other methods by then, we halved the current best estimates and assumed $\Omega_M = 0.3 \pm 0.02$ as a prior. Although this is not required (and indeed our analysis with current data does not make this assumption), it helps to reduce the error on $w$ significantly, and it certainly is sensible\footnote{When comparing our errors on the equation of state with those of the KAOS purple book, one should note that they fit simultaneously several parameters more, and that our fiducial model has a higher value of $w=-1/3$, which helps to reduce the errors further. A model with $w=-1$ would have larger errors in $w$.}. With these assumptions we conclude that we will be able to detect or rule out the mixing scenario at over 5$\sigma$ after marginalising over $w$. The estimated error bar on $\lambda$ is less than $0.1$ (and is degraded to about $0.13$ if no constraints on $\Omega_M$ are added). The 2d likelihood in the $\lambda-w$ plane is shown in the upper right hand corner of Figure (\ref{2daw}). 

Although we have assumed a constant $w$ here, the beauty of having both $d_L$ and $d_A$ information is that it allows us to separate the issue of mixing from the dynamics of the dark energy. At each redshift, axion-photon mixing should lead to fundamentally inconsistent values for $y(z)$ derived from $d_L$ and $d_A$ respectively. 

An estimate of cosmic parameters unbiased by mixing will be available from the Sloan Survey (Matsubara \& Szalay 2003) while a further test of mixing is provided by number counts versus redshift, $dN/dz$, which depends on $d_A(z)$. Since the volume of space as a function of redshift is very sensitive to $\Lambda$ number counts is a good test of acceleration. Generally, the number of objects in the range of affine parameter values $[y, y + \Delta y]$, is (e.g. Ellis 1971, Ribeiro \& Stoeger 2003)
\beq
dN = d_A^2 (1 + z)~n(y)~dy ~d\Omega
\eeq
where $n(y)$ is the number density of objects and $d\Omega$ is the differential solid angle at the observer. 
Axion-photon mixing alters galaxy number counts by reducing the apparent luminosity of objects at high redshift, at least in the visible range.  Since high redshift objects appear dimmer, the selection function $\psi$ is altered and faint galaxies will be lost. Therefore, there should be a deficit of objects relative to the case of no axion-photon mixing. 

If we therefore compare a standard $\Lambda$CDM model against a non-accelerating model with axion-photon mixing the difference in number counts at $z > 1$ is significant. One may consider variants of this basic idea such as the $dV/dz d\Omega$ test which, applied to the DEEP2 galaxy survey of $\sim 50k$ galaxies with redshifts $0.7 < z < 1.4$, should allow an estimate of $w$ today (unbiased by axion-photon mixing) to $\sim 10\%$ (Newman \& Davis, 2000). 

The mixing mechanism may be constrained in yet another manner however. Observations of the polarisation of light from gamma-ray burst (GRB) GRB021206 (Coburn and Boggs, 2003) have found linear polarisation levels of $\Pi = 0.80 \pm 0.2$, centered very near the maximum allowed by Compton scattering which strongly supports synchrotron radiation as the source of at least some GRB's. If GRB021206 is at a redshift $z > 0.1$ {\em and} Compton scattering is the source of the linear polarisation, then the near maximal value of $\Pi$ observed on earth leaves little room for depletion due to mixing. However, as pointed out in Cs\'aki \etal (2002), mixing is intrinsically inhomogeneous. It is possible to have certain lines of sight that experience essentially no mixing at all, depending on the magnetic field traversed. Hence, unless there is a high-z SNIa in the same narrow field of view as the GRB, a single event alone cannot rule out the mixing scenario. Further, the linear polarisation of the GRB may not be due to Compton scattering (Lazzati {\em et al}, 2003), in which case there might still be room for axion-photon mixing.    
\section{Conclusions}
The dimming of distant supernovae (SNIa) remains the most direct evidence for cosmic acceleration. Nevertheless  alternative explanations exist, such as axion-photon mixing in which roughly one third of all photons from distant SNIa are lost into axion states. We have pointed out that such mixing will not affect the angular-diameter distance $d_A(z)$ and hence will cause a fundamental asymmetry between measurements of the luminosity distance, $d_L(z)$, and $d_A(z)$ that can be searched for. 

In a first search for such asymmetry we have undertaken a joint analysis of high-redshift SNIa ($d_L(z)$) and radio galaxy data ($d_A(z)$). The results do not favour the loss of photons and hence disfavour mixing. Future data will improve the limits, and be able to test very generally for an asymmetry between $d_A(z)$ and $d_L(z)$. Number counts versus redshift are a promising test while estimates of $d_L(z)$ from SNAP and $d_A(z)$ from a large 2nd generation galaxy survey such as KAOS will allow axion-photon mixing to be detected or ruled out at more than $5\sigma$, almost independently of the dynamics of the dark energy, showing the power in constraining non-standard physics implicit in combining $d_L(z)$ and $d_A(z)$ data. 

\acknowledgments
We thank Steven Boggs, George Ellis, Takahiro Tanaka and John Terning for comments and/or stimulating  discussions. This research is supported by the JSPS/Royal Society and PPARC.




\begin{thebibliography}{}
\bibitem{snap} Aldering, G., \etal, 2002, SPIE, {\bf 4835}, 21
\bibitem{evidence} Barris, B.J., {\em et al}, 2003, {\em  Ap.J to appear}, astro-ph/0310843
\bibitem{martin} Bassett, B.A., and Kunz, M., 2003, astro-ph/0312443
\bibitem{rob} Boughn, S., and Crittenden, R., 2003, {\em to appear Nature}, astro-ph/0305001
\bibitem{blake} Blake, C., and Glazebrook, K., 2003, Astrophys.J. {\bf 594}, 665 
\bibitem{subir} Blanchard A., Douspis M., Rowan-Robinson M., Sarkar. S., 2003, astro-ph/0304237
\bibitem{AP} Cs\'aki, C., Kaloper, N., and Terning, J., 2002, Phys.\ Rev.\ Lett.\  {\bf 88}, 161302 (2002); ibid.
Phys.\ Lett.\ B {\bf 535}, 33
\bibitem{gzk} Cs\'aki, C., Kaloper, N., Peloso, M., and Terning, J.,
2003, JCAP {\bf 0305}, 005 
\bibitem{crit2} Christensson, M., and Fairbairn,M., 2003,
Phys.\ Lett.\ B {\bf 565}, 10 
\bibitem{GRB}Coburn W. and Boggs S.~E.,
2003, Nature {\bf 423}, 415 
\bibitem{bar_osc} Cooray, A., Hu, W., Huterer, D., and Joffre, M., 2001, Ap. J {\bf 557}, L7
\bibitem{daly}  Daly, R. A., \& Djorgovski, S. G.,  2003, astro-ph/0305197; 
\bibitem{d2} Daly, R.A., Guerra, E. J., 2002,  Astron.J. {\bf 124} 1831
\bibitem{crit1} Deffayet, C., Harari, D., Uzan, J.P., and Zaldarriaga, M., 2002,
Phys.\ Rev.\ D {\bf 66}, 043517 
\bibitem{ellis71} Ellis, G.F.R., 1971, in Proc. School "Enrico Fermi", Ed. R. K. Sachs, New York 
\bibitem{er} Erlich, J., and Grojean, C., 2002, Phys. Rev. D{\bf 65}, 123510, hep-ph/0111335
\bibitem[Etherington (1933)]{rec} Etherington, J.M.H., 1933, Phil. Mag. {\bf 15}, 761
\bibitem{corr3} Fosalba P., Gaztanaga E., Castander F., 2003, Astrophys. J. {\bf 597} L89 
\bibitem{gur} Gurvitis, L.I., 1994, Ap. J. {\bf 425}, 442 
\bibitem{jack} Jackson, J.C., 2003, astro-ph/0309390.
\bibitem{jack2} Jackson, J.C., and Dodgson, M.,  1997, MNRAS, {\bf 285} 806 
\bibitem{kaos} The KAOS purple book, http://www.noao.edu/kaos/KAOS\_Final.pdf, 2003
\bibitem{sn1a} Knop, R.A., \etal, 2003, {\em Ap.J to appear}, astro-ph/0309368
\bibitem{grbalt} Lazzati, D., Rossi, E., Ghisellini, G., and Rees, M.~J., 2003, 
arXiv:astro-ph/0309038.
\bibitem{da} Lima, J.~A., and Alcaniz,J.~S., 2002,
Astrophys.\ J.\  {\bf 566}, 15 
\bibitem{linder} Linder, E. V.,  2003, Phys.Rev. D{\bf 68} 083504
\bibitem{MR} Mamon, G.A., and Roukema, B.F.,
arXiv:astro-ph/0212169 ; Roukema,B. F., Mamon,G. A., Bajtlik,S., (2002)  
A\&A {\bf 382}, 39  
\bibitem{ms} Matsubara, T., \& Szalay, A.S., 2002, Ap. J., {\bf 574}, 1 
\bibitem{gb1} Mortsell, E., Bergstrom, L., Goobar, A., 2002, Phys.Rev. D{\bf 66}, 047702
\bibitem{gb2} Mortsell, E., and Goobar, A., 2003, JCAP 0304, 003
\bibitem{ND} Newman, J.A., and Davis, M., 2000, Ap. J, {\bf 534} L11; astro-ph/9912366  
\bibitem{wmap} Nolta, M.R., \etal, 2003, astro-ph/0305097
\bibitem{2df} Outram, P.J., {\em et al}, 2003, MNRAS {\em to appear}, astro-ph/0310873
\bibitem{stoeger} Ribeiro, M., and Stoeger,W., 2003, Ap. J. {\bf 592}, 1  
\bibitem{SEF} Schneider, P., Ehlers, J., and Falco, E.E., 1992, Gravitational Lenses, Springer-Verlag, Berlin.
\bibitem{sdss} Scranton, R., \etal, 2003, astro-ph/0307335
\bibitem{future} Seo, H-J, \& Eisenstein, D.J., astro-ph/0307460 (2003)
\end{thebibliography}
\end{document}